\documentclass[showpacs,twocolumn,aps]{revtex4}
\usepackage{amssymb}
\usepackage{amsmath}
\usepackage{graphicx}
\usepackage{lscape}
\usepackage{booktabs}
\usepackage[colorlinks,linkcolor=blue,citecolor=blue]{hyperref}
\begin{document}
\title{Energy dependence of light (anti)nuclei and (anti)hypertriton production\\ in the Au-Au collision from $\sqrt{s_{\rm{NN}}} =5.0$ to $5020$ GeV}

\author{Zi-Jian Dong$^{1}$, Gang Chen$^{1}\footnote{chengang1@cug.edu.cn}$, Quan-Yu Wang$^{1}\footnote{wangqy@cug.edu.cn}$, Zhi-Lei She$^1$, Yu-Liang Yan$^{2}$,\\
 Feng-Xian Liu$^1$, Dai-Mei Zhou$^3$ and Ben-Hao Sa$^{2,3}$}

\affiliation{%
 1 School of Mathematics and Physics;Network and Educational Technology Center, China University
of Geoscience, Wuhan 430074, China.\\
2 China Institute of Atomic Energy, P.O. Box 275 (10), Beijing 102413, China.\\
3 Key Laboratory of Quark and Lepton Physics (MOE) and Institute of Particle
Physics, Central China Normal University, Wuhan 430079, China.}

\begin{abstract}
The energy dependence of light (anti)nuclei and (anti)hypertriton production are investigated in central Au-Au collisions from AGS up to LHC energies at midrapidity, using the parton and hadron cascade model (PACIAE) together with the dynamically constrained phase-space coalescence model(DCPC). We find that the yields, yield ratios of the antiparticles to their corresponding particles, the coalescence parameters $B_A$ and the strangeness population factor $s_3$ of light (anti)nuclei and (anti)hypertriton strongly depend on the energy. Furthermore, we analyze and discuss the strangeness population factor $s_3$ and the coalescence parameters $B_A$, and find a transition point near by 20 GeV. These results thus suggest the potential usefulness of the $s_3$ and $B_A$ of light nuclei production in relativistic heavy-ion collisions as a direct probe of the transition point associated with the QCD critical phenomena. The results from PACIAE+DCPC model are well consistent with experimental data.
\end{abstract}
\pacs{25.75.-q, 24.85.+p, 24.10.Lx}

\maketitle

\section{Introduction}
In recent decades, the problem of antimatter has attracted considerable attention in particle and nuclear physics, astrophysics, cosmology and other fields of modern physics, since the first antiparticle(positron)~\cite{1} was discovered by Anderson. The antiprotons~\cite{2} and antineutrons~\cite{3} were discovered by Wiegand and Piccioni in 1955 and 1956, respectively. Then some light anti-nuclei, such as anti-deuteron, anti-triton, anti-helium-3, anti-helium and anti-hypernucleus, were also produced and detected in experiments~\cite{4,5,6,7,8,9}. The high temperature and high baryon density condition similar to the initial stages of the big bang, created by heavy ion collision experiments, is suitable for the study of light (anti)nuclei production and its energy evolution.

The PHENIX and STAR Collaboration have reported their light nuclei production data~\cite{10,11,12,13} for Au-Au collisions at $\sqrt{s_{\rm{NN}}} = 7.7-200$ GeV. And ALICE has also published some (anti)hadrons and light (anti)nuclei production of pp~\cite{14,15} collisions at $\sqrt{s_{\rm{NN}}} = 0.9,2.76$ and $7$ TeV , as well as  Pb-Pb~\cite{16,17,18} collisions at $\sqrt{s_{\rm{NN}}} = 6.3-17.5$ GeV and 2.76 TeV.

On the other hand, the theoretical calculation of light (anti)nuclei is usually divided into two steps. Firstly, one can obtain the nucleon and hyperon yield in some selected models such as the transport model. On the second step, the light nuclei (anti-nuclei) are studied by the reasonable hadron final-state coalescence models~\cite{19,20,21,22,23}, such as the phase-space coalescence model~\cite{24,25,26} and/or the statistical model~\cite{27,28}, etc. For example, the production of light nuclei (hyper-nuclei) in the Au-Au and Pb-Pb collisions at relativistic energies was described theoretically by the coalescence + blast-wave method~\cite{29,30,31} or the coalescence + a multiphase transport (AMPT) model~\cite{32} and the hybrid UrQMD model~\cite{33}, respectively.

In this work, we will take an approach based on the parton and hadron cascade model (PACIAE)~\cite{34} + the dynamically constrained phase-space coalescence model (DCPC)~\cite{35}. Which has been used to predict the light nuclei (antinuclei) yield in transverse momentum and rapidity space for non-single diffractive proton-proton collisions at $\sqrt{s_{\rm{NN}}}$ = 7 TeV~\cite{35}, and the production of light nuclei and hypernuclei in Au-Au collisions at $\sqrt{s_{\rm{NN}}}$ = 200 GeV~\cite{36,37,38}, as well as Pb-Pb collisions at $\sqrt{s_{\rm{NN}}}$ = 2.76 TeV~\cite{39}. Moreover, this method has been applied to explore the energy dependence of the antiparticle to particle ratio in high energy proton-proton collisions~\cite{40}. In this paper, we will extend this method to investigate the productions of light (anti)nuclei and (anti)hypernuclei in the 0-10\% centrality Au-Au collisions at $\sqrt{s_{\rm{NN}}} = 5,6,7,7.7,9.9,11.5,14.5,19.6,27,39,62.4,130,200,800$, $2760,5020 $ GeV, respectively.

This paper is organized as follows: in Sec. II, we briefly introduce the PACIAE and DCPC models. The results of our simulated light (anti)nuclei and (anti)hypernuclei yields and their ratios will be provided to compare with experimental data in Sec. III. In the end, we will offer a brief summary in Sec.IV.

\section{MODELS}
The PYTHIA model (PYTHIA 6.4~\cite{41}) is devised for high energy hadron-hadron (hh) collisions. In this model, a hh collision is decomposed into parton-parton collisions described by the leading order perturbative QCD (LO-pQCD). For the soft interaction, a nonperturbative process is considered empirically. The initial- and final-state QCD radiations and the multiparton interactions are also taken into account. Therefore, the consequence of a hh collision is a partonic multijet state composed of (anti)diquarks, (anti)quarks, and gluons, as well as a few hadronic remnants. This is then followed by the string construction and fragmentation. A hadronic final state is obtained for a hh collision eventually.

The parton and hadron cascade model PACIAE~\cite{34} is based on PYTHIA 6.4 and is devised mainly for the nucleus-nucleus collisions. In the PACIAE model, first, the nucleus-nucleus collision is decomposed into separate nucleon-nucleon (NN) collisions according to the collision geometry and NN total cross section. Each NN collision is described by the PYTHIA model with the string fragmentation switches off and the (anti)diquarks randomly break into (anti)quarks. So the consequence of one NN collision is a partonic initial state composed of quarks, antiquarks and gluons. This partonic initial state is regarded as the quark-gluon matter (QGM) formed in relativistic nucleus-nucleus collisions. Second, the parton rescattering proceeds by the 2$\rightarrow$2 LO-pQCD parton-parton scattering~\cite{42}. In addition, a K factor is introduced here to account for higher order and nonperturbative corrections. Third, hadronization happens after parton rescattering. The partonic matter can be hadronized by the Lund string fragmentation scheme~\cite{41} and/or the phenomenological coalescence model~\cite{40}. Finally, the hadronic matter continues rescattering until the hadronic freeze-out (the exhaustion of the hadron-hadron collisions). We refer to~\cite{40} for the details.

With the final state particles generated by the PACIAE model, we can then calculate the production of light (anti)nuclei with the DCPC model. In quantum statistical mechanics~\cite{43}, one cannot precisely define both position $\vec q\equiv (x,y,z)$ and momentum $\vec p\equiv (p_x,p_y,p_z)$ of a particle in the six-dimension phase-space because of the uncertainty principle $\Delta\vec q\Delta\vec p \geqslant h^3$. We can only say that this particle lies somewhere within a six-dimension quantum box or state with a volume of $\Delta\vec q\Delta\vec p$. A particle state occupies a volume of $h^3$ in the six-dimension phase-space~\cite{43}. Therefore, one can estimate the yield of a single particle by defining an integral
\begin{equation}
Y_1=\int_{H\leqslant E} \frac{d\vec qd\vec p}{h^3},
\end{equation}
where $H$ and $E$ are the Hamiltonian and energy of the particle, respectively. Similarly, the yield of the N particle cluster can be estimated as the following integral:
\begin{equation}
Y_N=\int ...\int_{H\leqslant E} \frac{d\vec q_1d\vec p_1...d\vec q_Nd\vec p_N}{h^{3N}}.
\end{equation}
In addition, Eq. (2) must satisfy the following constraint conditions:
\begin{equation}
m_0\leqslant m_{inv}\leqslant m_0+\Delta m,
\end{equation}
\begin{equation}
 q_{ij}\leqslant D_0,(i\neq j;i,j= 1,2,\cdots,N).
\end{equation}
Where
\begin{equation}
m_{inv}= \sqrt{(\sum^N_{i= 1}E_i)^{2}-(\sum^N_{i= 1}p_i)^{2}},
\end{equation}
and $E_i,p_i\hspace{0.2cm}(i=1,2,\cdots ,N)$ are the energies and momenta of particles, respectively. $m_0$ and $D_0$ stand for, respectively, the rest mass and diameter of light (anti)nuclei, $\Delta m$ refers to the allowed mass uncertainty, and $q_{ij}=|q_i-q_j|$ is the vector distance between particles $i$ and $j$. Because the hadron position and momentum distributions from transport model simulations are discrete, the integral over continuous distributions in Eq.(2) should be replaced by the sum over discrete distributions in the phase-space.

\section{Results and Discussion}
\begin{figure}[htbp]
\includegraphics[width=8.8cm]{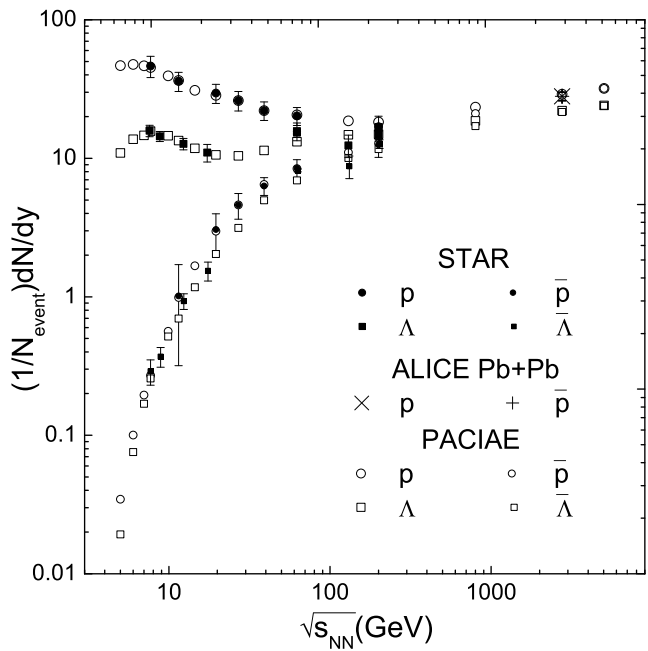}
\caption{The integrated yield dN/dy distributions of particles $p,\overline{p},\Lambda,\overline{\Lambda}$ in (the centrality of 0-10\% for $p,\overline{p}$  and 0-5\% for $\Lambda,\overline{\Lambda}$) Au-Au collisions at midrapidity $|y|<0.5$, as a function of energy. The open symbols represent PACIAE model results and the solid symbols are the data points from experimental collaboration~\cite{16,44,45}.}
\label{tu1}
\end{figure}

Firstly, we produce the final state particles using the PACIAE model. In the PACIAE simulations we assume that hyperons are heavier than $\Lambda$  decay already. The model parameters are fixed on the default values given in PYTHIA~\cite{41}. However, the K factor as well as the parameters parj(1), parj(2), and parj(3), which are relevant to the hadrons production in PYTHIA, are tuned by fitting the STAR data of $p,\overline{p},\Lambda$ and $\overline{\Lambda}$ in Au+Au collisions at $\sqrt{s_{\rm{NN}}} = 7.7-200$ GeV in 0-10\% centrality~\cite{44,45}. Specific details of this method is similar to the paper~\cite{37,40}. We have generated more than 10 million events by the PACIAE model for the 0-10\% centrality Au-Au collisions at $\sqrt{s_{\rm{NN}}} = 5.0 - 5020$ GeV with $|y| < 0.5$, $0 < p_T < 5$ GeV/c. Fig \ref{tu1} shows that the integrated yield dN/dy distributions of $p,\overline{p},\Lambda$ and $\overline{\Lambda}$ as a function of the energy. The results of the model with open symbols are consistent with the STAR and ALICE data( solid symbols), which also has been used to determine the input parameters. And the (anti)proton yields of Pb-Pb collisions are used to compare with those of Au-Au collisions at $\sqrt{s_{\rm{NN}}} = 2.76$ TeV when their $N_{part}$~\cite{10,46,47} are similar.

\begin{figure}[htbp]
\includegraphics[width=8.5cm]{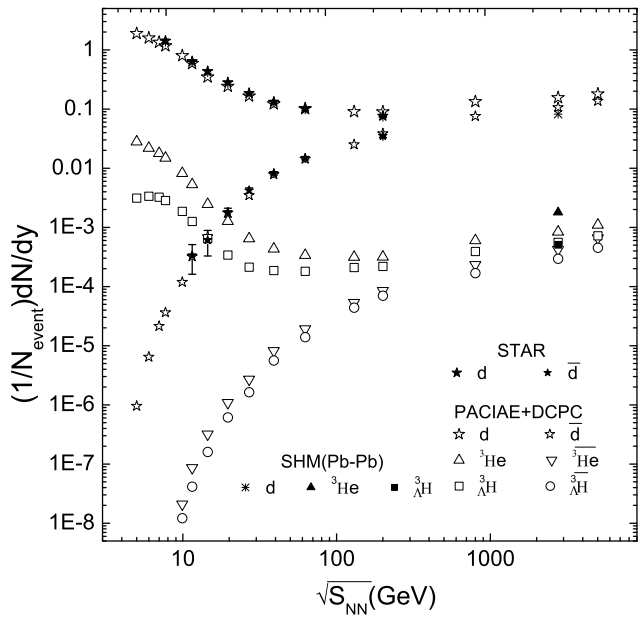}
\caption{The dN/dy distributions of $d$, $\overline{d}$, $^3{He}$, $\overline{^3{He}}$, $_{\Lambda}^3H$ and $\overline{_{\overline\Lambda}^3H}$, produced in Au-Au collisions with centrality 0-10\% and midrapidity, as a function of energy. The solid symbols represent PACIAE model results and the open symbols are the data points from STAR collaboration~\cite{9,44,48} or SHM model~\cite{49}.}
\label{tu2}
\end{figure}

Then, we show that predictions using the PACIAE+DCPC model can be used to describe quantitatively the measured energy dependence of light (anti)nuclei $d$($\overline d$), $^3{He}$($\overline{^3{He}}$), as well as $_{\Lambda}^3H$($\overline{_{\overline\Lambda}^3H}$) integrated yields dN/dy over a very wide energy range from $\sqrt{s_{\rm{NN}}} = 5.0$ GeV to $5.02$ TeV, as shown in Fig~\ref{tu2}. The calculations, here and in the following, are performed in the midrapidity region $|y|<0.5$ for 0-10\% centrality Au-Au collisions.
From Fig.2, we can see that, in low-energy regions, the yields dN/dy of light nuclei ($d, {^3{He}}$) and hypertritons (${_{\Lambda}^3H}$) decrease with the increase of energy, whereas the yield of their antiparticles are increasing by contrast. In the end, the yields of light nuclei and their corresponding antinuclei converge to a unified value when the energy is greater than 200 GeV.
 It is observed that the results from our simulation agree very well with the measurements at STAR ~\cite{9,44} for the $d, \overline d$) yields. We also compare results of DCPC model with the SHM models~\cite{49} under their approximate $N_{part}$ condition in Pb-Pb collisions.

\begin{figure}[htbp]
\includegraphics[width=8.5cm]{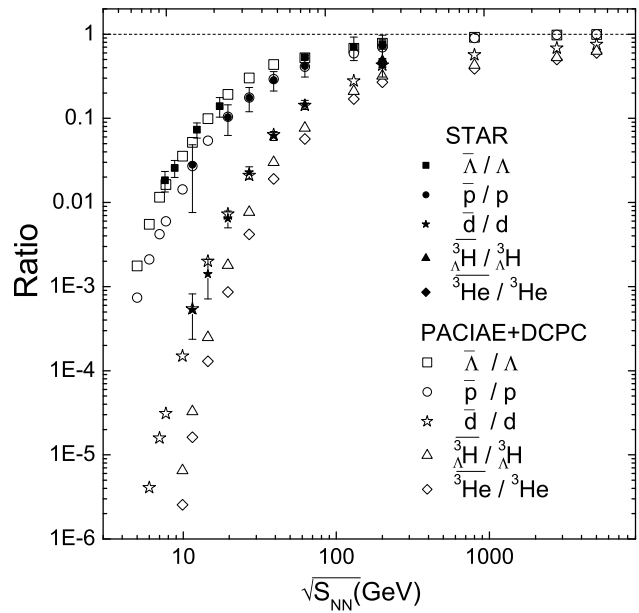}
\caption{The yield ratios distributions of $\overline p/p$, $\overline \Lambda/\Lambda$, $\overline d/d$, $\overline{^3{He}}/^3{He}$, $\overline{_{\overline\Lambda}^3H}/_{\Lambda}^3H$ in the midrapidity region $|y|<0.5$ and the the 0-10\% centrality Au-Au collisions, as a function of energy. The solid symbols represent DCPC model results and the open symbols are the data points from STAR collaboration~\cite{8,16,44}.}
\label{tu3}
\end{figure}

In Fig~\ref{tu3} we present the energy dependence of $\overline p/p$, $\overline \Lambda/\Lambda$, $\overline d/d$, $\overline{^3{He}}/^3{He}$, $\overline{_{\overline\Lambda}^3H}/_{\Lambda}^3H$ yield ratios in the $0-10$\% centrality Au-Au collisions at midrapidity $|y|<0.5$. We can see that, in a low energy range, the anti-particle to particle ratios for nucleon, hyperon, light nuclei and hypertriton all increase rapidly with the growth of energy; in a high energy range, their ratio gradually becomes saturated asymptotically equal to 1. To facilitate the comparison, the experimental results from STAR~\cite{8,16,44} are also drawn in the figure with the solid points. It is clear that the results of the PACIAE+DCPC model are in agreement with the experiments.

In heavy-ion collisions, the coalescence process of light (anti)nuclei, and (anti)hypernuclei are historically described~\cite{21,22,23} by the coalescence parameter $B_A$, which is used to describe the difficulty of synthesizing nucleus. The differential invariant yield is related to the primordial yields of nucleons and is described by the equation
\begin{equation}
E_A\frac{d^3N_A}{d^3p_A} \approx B_A\Big{(}E_p\frac{d^3N_p}{d^3p_p}\Big{)}^{A},
\end{equation}
where $Ed^3N/d^3p$ stands for the invariant yields of nucleons or light (anti)nuclei, and (anti)hypernuclei, and A is the atomic mass number. $p_A$, $p_p$ denote their momentum, with $p_A$ = A$p_p$ assumed.

\begin{figure}[htbp]
\includegraphics[width=8.5cm]{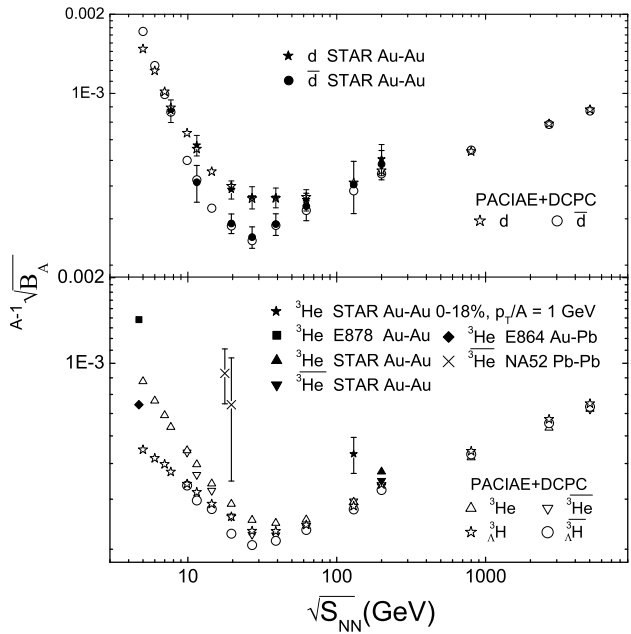}
\caption{Coalescence parameters $B_A$ as a function of energy for light (anti)nuclei ($p_T/A=0.65$ GeV/c) in Au-Au collisions at midrapidity $|y|<0.5$. The open symbols represent model results evaluated for different types of (anti)nuclei. The solid points are the data points from experimental measurements evaluated with (anti)deuterons,(anti) hypertriton and (anti)helium-3~\cite{44,50}.}
\label{tu4}
\end{figure}

The phase-space information of $p(\overline p),\Lambda(\overline\Lambda)$ which is used as an input for the coalescence prescription, can be reproduced by the PACIAE model. The coalescence parameters $B_A$ can be evaluated by comparing the integrated yields of the light (anti)nuclei and the primordial (anti)nucleons. Figure~\ref{tu4} presents DCPC model results of $\sqrt[A-1]{B_A}$ as a function of energy(open points), compared with experimental data of Au-Au~\cite{44} and Pb-Pb~\cite{50} collisions(solid points), respectively. The distributions of $\sqrt[A-1]{B_A}$ show a minimum value around $\sqrt{s_{\rm{NN}}} =20$ GeV.  Below 20 GeV, the coalescence parameter $\sqrt[A-1]{B_A}$ decreases as increasing energy, implying that the emitting source size increases with the growing collision energy. These non-monotonic patterns are consistent with the minimum value observed for the energy dependence of the viscous coefficients and $\pi$ HBT results~\cite{51}. However, the $\sqrt[A-1]{B_A}$ is gradually increasing in the relative high energy region. $\sqrt[A-1]{B_A}$ of antiparticles increase from less than that of their particles to approximately equal. The coalescence parameter $\sqrt[A-1]{B_A}$ of Pb-Pb collisions is larger than that of Au-Au collisions in the same energy. All these imply the coalescence parameter $B_A$ is related to density of corresponding components. In addition, $\sqrt[A-1]{B_A}$ of particles and antiparticles are almost the same at high collision energy, indicating that the emitted source size of particles is approximately equal to that of their anti-particles at the same high collision energy.

\begin{figure}[htbp]
\includegraphics[width=8.5cm]{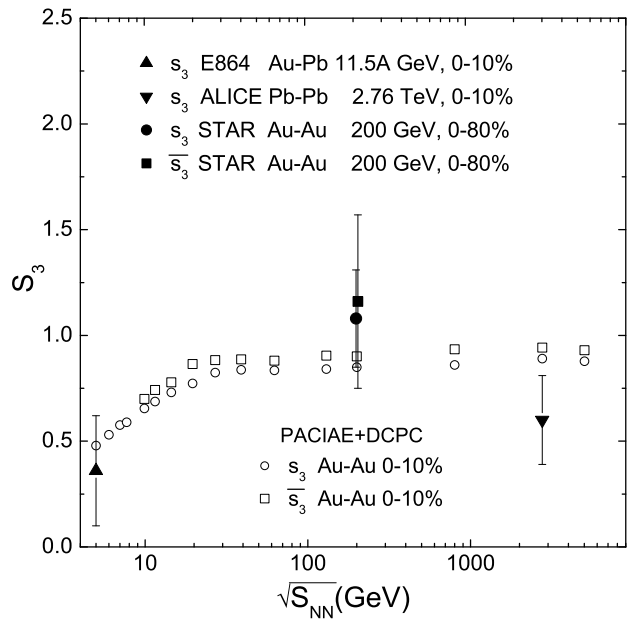}
\caption{The distributions of strangeness population factor $s_3$ and $\overline{s_3}$ as a function of energy. The open symbols represent our model results and the solid symbols are the data points from experimental collaboration~\cite{9,52,53}.}
\label{tu5}
\end{figure}

In order to compare (anti)nuclei and (anti)hypernuclei, we also study the strangeness population factor~\cite{9,52,53}
\begin{equation}
s_3 = (_{\Lambda}^3H\times p)/(^3{He}\times \Lambda)£¬
\end{equation}
\begin{equation}
\overline{s_3} = (\overline{_{\overline\Lambda}^3H}\times \overline p)/(\overline{^3{He}}\times \overline\Lambda).
\end{equation}

 In Fig~\ref{tu5}, we show the energy dependence of the strangeness population factor $s_3$ and $\overline{s_3}$, which increase with the growth of energy below 20 GeV, and then saturate around 0.8 in the high energy region.

To get a better sense of what is happening, $s_3$ is represented as $\frac{_{\Lambda}^3H}{\Lambda \times p \times n}$ = $s_3$$\frac{^3{He}}{p \times p \times n}$, where $\frac{_{\Lambda}^3H}{\Lambda \times p \times n}$ and $\frac{^3{He}}{p\times p\times n}$ respectively stand for the degree of difficulty of $_{\Lambda}^3H$ and $^3{He}$ generation, which is approximately equal to $B_A$~\cite{21,22,23}. $s_3$ gradually increase as the collision energy grows, indicating that relative to ordinary nuclei, the higher the energy is, the easier it is to generate the hypernuclei. And above 20 GeV, $s_3$ is independent of energy and approximately equal to 0.8, implying that $_{\Lambda}^3H$ generation is harder than $^3{He}$. It is interesting that we find at the same energy, $\frac{_{\Lambda}^3H}{\Lambda \times p \times n} \neq \frac{\overline{_{\overline\Lambda}^3H}}{\overline\Lambda \times \overline p \times \overline n}$ and $\frac{^3{He}}{p\times p\times n} \neq \frac{\overline{^3{He}}}{\overline p\times \overline p\times \overline n}$,  that is to say the probability of $_{\Lambda}^3H$ and $^3{He}$ generation is different from that of their antimatter, respectively. $s_3$ and $\overline {s_3}$, however, are approximately equal, indicating that strange to nonstrange light nuclei relative production ratio are the same for particle and antiparticle.

At the same energy, there are three factors relevant for the(hyper)nuclei generation, i.e., properties of corresponding component, density  of corresponding component and the emitted source size~\cite{54} of difference between particles and their anti-particles. We assume the emitted source size of particles is the same as that of their anti-particles. On the other hand, the phase space density of p and Lambda generated by PACIAE is similar to their antiparticle sector. It is natural to observe that the strange population factor $s_3$ shows no difference from $\overline {s_3}$.

\section{Conclusion}
In this paper we use the PACIAE+DCPC model to investigate the energy dependence of light (anti)nuclei and (anti)hypertriton production in the 0-10\% centrality Au+Au collisions at midrapidity $|y| < 0.5$ and $0<p_t<5$ GeV/c. Firstly, the integrated yields dN/dy of $p$, $\overline p$, $\Lambda$ and $\overline \Lambda$ are calculated by the PACIAE model for a wide range of collision energy. The result shows that the yields of the antinucleon or antihyperons increase as the energy increases, whereas yields of Nucleon or hyperon decrease first and then increase. Secondly, the integrated yields dN/dy of $d(\overline d)$, $^3{He}(\overline{^3{He}})$, $_{\Lambda}^3H(\overline{_{\overline\Lambda}^3H})$ are also calculated by the DCPC model. Their distributions are similar to those of nucleon or hyperon. Then we find the yield ratios of light anti-nuclei to light nuclei and anti-hypertriton to hyper-triton all increase asymptotically to 1 with the growth of energy. Furthermore, we also discuss coalescence parameters $B_A$ of light (anti)nuclei and (anti)hypernuclei. We find the emitting source size is increasing with collision energy while the coalescence paramter $B_A$ is correlated with the density of corresponding constituent particles. At last, we calculate the distributions of strangeness population factor $s_3(\overline{s_3})$ as a function of energy. It can be concluded in this study that $_{\Lambda}^3H$ is generally harder to generate than $^3{He}$ while the probability for the hypernuclei formation is increasing with energy below 20 GeV.

It is worth noting that we found a transition point exist at 20~GeV, through extracted the collision energy dependence of $B_A$ and $s_3(\overline{s_3})$, suggesting that the critical endpoint in the QCD phase diagram may have been reached or closely approached in these collisions. This question needs to be further studied.

Our results are well consistent with experimental data.
The consistency between the PACIAE+DCPC results and the corresponding experimental data demonstrates that the PACIAE+DCPC method is able to describe the production of light (anti)nuclei and (anti)hypernuclei for extensive range of heavy-ion collision energy.

\begin{center} {ACKNOWLEDGMENT} \end{center}
Finally, we acknowledge the financial support from NSFC (11475149).
The authors thank PH.D. Liang Zheng for helpful discussions. The authors thank PH.D. Yi-long Xie for improving the English.


\begin{thebibliography}{99}
\bibitem{1} C. D. Anderson, Phys. Rev. 43, 491 (1933).
\bibitem{2} O. Chamberlain, E. Segr\`e, and C. Wiegand, Phys. Rev. 100. 947 (1955).
\bibitem{3} B. Cork, G.R. Lambertson, O. Piccioni et al., Phys. Rev.104, 1193 (1956).
\bibitem{4} D.E. Dorfan, J. Eades, L.M. Lederman et al., Phys. Rev. Lett. 14, 1003 (1965).
\bibitem{5} Y. M. Antipov et al., Yad. Fiz. 12, 311 (1970).
\bibitem{6} N. K. Vishnevsky et al., Yad. Fiz. 20, 694 (1974).
\bibitem{7} M. Danysz and J. Pniewski, Philos. Mag. Ser. 7 44, 348 (1953).
\bibitem{8} H. Agakishiev et al. (STAR Collaboration), Nature 473, 353 (2011).
\bibitem{9} B. I.Abelevet al. (STARCollaboration), Science 328, 58 (2010).
\bibitem{10} A. Adare et al. (PHENIX Collaboration), Phys. Rev. C 93, 024901 (2016).
\bibitem{11} STAR Collaboration Studying the Phase Diagram of QCD Matter at RHIC (2014).
\bibitem{12} STAR Collaboration, light (anti)nucleus prodution in $\sqrt{s_{\rm{NN}}}$ = 7.7-200 GeV Au-Au collisions in the STAR Experiment (2012).
\bibitem{13} STAR Collaboration, arXiv:1710.00436v1 [nucl-ex] (2017).
\bibitem{14} S.Natasha(for the ALICE Collaboration) J. Phys. G: Nucl. Part. Phys.38, 124189 (2011).
\bibitem{15} ALICE Collaboration arXiv:1709.08522v1 [nucl-ex] (2017).
\bibitem{16} B. Abelev et al. (ALICE Collaboration) Phys. Rev. C 88,044910 (2013).
\bibitem{17} T. Anticic et al. (NA49 Collaboration) Phys. Rev. C 94, 044906 (2016).
\bibitem{18} ALICE Collaboration arXiv:1710.07531v1 [nucl-ex] (2017).
\bibitem{19} S.T. Butler, C.A. Pearson, Phys. Rev. 129, 836 (1963).
\bibitem{20} A. Schwarzschild, C. Zupancic, Phys. Rev. 129, 854 (1963).
\bibitem{21} H.H. Gutbrod, A. Sandoval, P.J. Johansen et al., Phys.Rev. Lett. 37, 667 (1976).
\bibitem{22} J. Gosset, H.H. Gutbrod, W.G. Meyer et al., Phys. Rev.C 16, 629 (1977).
\bibitem{23} M.C. Lemaire, S. Nagamiya, S. Schnetzer et al., Phys. Lett.B 85, 38 (1979).
\bibitem{24} R. Mattiello, H. Sorge, H. St¡§ocker et al., Phys. Rev. C 55,1443 (1997).
\bibitem{25} L.W. Chen, C.M. Ko, Phys. Rev. C 73, 044903 (2006).
\bibitem{26} S. Zhang, J.H. Chen, H. Crawford et al., Phys. Lett. B 684, 224 (2010).
\bibitem{27} V. Topor Pop, S. Das Gupta, Phys. Rev. C 81, 054911 (2010).
\bibitem{28} A. Andronic, P. Braun-Munzinger, J. Stachel et al., Phys.Lett. B 697, 203 (2011).
\bibitem{29} L. Xue, Y.G. Ma, J.H. Chen et al., Phys. Rev.C85, 064912(2012).
\bibitem{30} C.S. Zhou, Y.G. Ma, S. Zhang, arXiv:1501.04386v1 [nuclth] (2015).
\bibitem{31} N. Shah, Y.G. Ma, J.H. Chen et al., Phys. Lett. B 754, 6 (2016).
\bibitem{32} L.L. Zhu, C.M. Ko, X.J. Yin, arXiv:1510.03568v1 [nucl-th] (2015).
\bibitem{33} J. Steinheimer, K. Gudima, A. Botvina et al., Phys. Lett.B 714, 85 (2012).
\bibitem{34} B.H. Sa, D.M. Zhou, Y.L. Yan et al., Comput. Phys. Commun.183, 333 (2012).
\bibitem{35} Y.L. Yan, G. Chen, X.M. Li et al.,Phys. Rev. C 85, 024907 (2012).
\bibitem{36} G. Chen, Y.L. Yan, D.S. Li et al., Phys. Rev. C 86, 054910 (2012).
\bibitem{37} G. Chen, H. Chen, J. Wu et al., Phys. Rev. C 88, 034908 (2013).
\bibitem{38} G. Chen, H. Chen, J.L. Wang et al., J. Phys. G: Nucl.Part. Phys. 41, 115102 (2014).
\bibitem{39} Z. L. She, G. Chen, et al., Eur. Phys. J. A (2016) 52: 93.
\bibitem{40} J.L. Wang, D.K. Li, H.J. Li et al., Int. J. Mod. Phys. E 23, 1450088 (2014).
\bibitem{41} T. Sj¡§ostrand, S. Mrenna, and P. Skands, J. High Energy Phys.05 (2006) 026.
\bibitem{42} B. L. Combridge, J. Kripfgang, and J. Ranft, Phys. Lett. B 70,234 (1977).
\bibitem{43} K. Stowe, An Introduction to Thermodynamics and Statistical Mechanics (Cambridge University, New York, 2007) and R. Kubo, Statistical Mechanics (North-Holland, Amsterdam, 1965).
\bibitem{44} Ning Yu(for the STAR Collaboration) Nuc. Phys. A 967 (2017) 788.
\bibitem{45} M. M. Aggarwal et al. (STAR Collaboration) Phys. Rev. C 83, 024901 (2011).
\bibitem{46} B. Abelev et al. (ALICE Collaboration) Phys. Rev. C 88, 044909 (2013).
\bibitem{47} B. Alver et al. (PHOBOS Collaboration) Phys. Rev. C 94, 024903 (2016).
\bibitem{48} J. Beringer et al. (Particle Data Group) Phys. Rev. D 86, 010001 (2012).
\bibitem{49} P. Michal and L. Jean and P. Vojt et al.,Phys. Rev. C 88, 034907 (2013).
\bibitem{50} C. Adler et al. (STAR Collaboration) Phys. Rev. Lett. 87, 262301 (2001).
\bibitem{51} PHENIX Collaboration arXiv:1410.2559v1 [nucl-ex] (2014).
\bibitem{52} K. J. Sun and L. W. Chen Phys. Rev. C 95, 044905 (2017).
\bibitem{53} ALICE Collaboration arXiv:1506.08453v2 [nucl-ex] (2016).
\bibitem{54} A. Adare et al. (PHENIX Collaboration), arXiv:1410.2559 [nucl-ex] (2014).

\end{thebibliography}
\end{document}